\documentclass[intlimits,twoside,a4paper]{article}

\usepackage{amsmath,amssymb}
\usepackage{graphicx}
\usepackage{txfonts}
\usepackage[eqsecnum]{cmpj2}
\articletype{Regular article}

\title{Strange Inter-layer Properties of Ba(Fe$_{1-x}$Co$_{x}$)$_{2}$As$_{2}$ Appearing in Ultrasonic Measurements}

\author{S. Simayi$^{*}$\refaddr{label1}, K. Sakano\refaddr{label1}, H. Takezawa\refaddr{label1}, M. Nakamura\refaddr{label1}, Y. Nakanishi\refaddr{label1}, K. Kihou\refaddr{label2}, M. Nakajima\refaddr{label2}, C. Lee\refaddr{label2}, A. Iyo\refaddr{label2}, H. Eisaki\refaddr{label2}, S. Uchida\refaddr{label3}, M. Yoshizawa\refaddr{label1}}
      
\addresses{
\addr{label1} Graduate School of Engineering, Iwate University, Morioka 020-8551, Japan
\addr{label2} National Institute of Advanced Industrial Science and Technology (AIST), Tsukuba 305-8568, Japan
\addr{label3} Graduate School of Science, The University of Tokyo, Tokyo 113-0033, Japan
}

\begin{document}

\maketitle

\begin{abstract}
We have investigated the elastic constant $C_{\rm 33}$ of Ba(Fe$_{1-x}$Co$_{x}$)$_{2}$As$_{2}$ with eight different Co concentrations by ultrasonic measurement. We found remarkable elastic anomalies near the quantum critical point. We have studied them by measuring the electrical resistivity, heat capacity, and ultrasonic attenuation in addition to the elastic constant. These results have revealed that the inter-layer three-dimensional properties appearing in $C_{\rm 33}$ to be possibly originated from the magnetic character of these materials. Our data about the elastic constant $C_{\rm 33}$ highlight the importance of controlling the $c-$axis length in the emergence of superconductivity in iron-based superconductors.

\keywords iron-based superconductor, elastic constant, lattice fluctuation
\end{abstract}

\section{Introduction}
The discovery of superconductivity in LaFeAsO$_{1-x}$F$_{x}$ with a transition temperature of $T_{\rm sc}$ = 26 K by Hosono group in 2008,\cite{Hosono} puts iron pnictide superconductors in the first stage of experimental and theoretical studies in superconducting research. 
The discovery of iron-based superconductors with a high phase transition temperature was unexpected, because they contain a magnetic element Fe. For this reason they have opened a new avenue of research.
BaFe$_{2}$As$_{2}$ undergoes a structural phase transition from tetragonal to orthorhombic at 140 K. 
By doping of Co ion, magnetic and structural phase transitions separate from each other and superconducting phase gradually appears. 
Within very short time, the critical temperature of iron based superconductors has been increased up to 55 K by replacing La with other rare earth elements.\cite{Ren}
To date, various structural classes of iron-based superconductors have been found. All share a common structural feature of having a tetrahedral structure and iron atoms tetrahedrally coordinated with pnictogen (P, As) or chalcogen (S, Se, Te) elements, and that iron atoms are located in the center of the tetrahedron. These tetrahedrally coordinated slabs provide a quasi-two-dimensional character to the crystal since they form atomic bonds with the FeAs layer. 

The same as that in cuprate superconductors, the pairing mechanism in what is still under debate. 
For all iron-based superconductors, electron-phonon coupling cannot explain the superconducting properties,\cite{Boeri} so the pairing mechanism has been focusing on other parameters such as the role of magnetism and orbitals. 
For instance, at the beginning it was proposed that the superconducting gap functions mediated by the spin fluctuation.\cite{Mazin,Kuroki} 
Christianson et al. supported this theory with their experimental results.\cite{Christ} 
However, Sato et al. and Lee et al. rejected this theory with their experimental results on the impurity effect for several iron-based superconductors.\cite{Sato, Satomi} 
On the other hand, Kontani et al. and Yanagi et al. proposed a conventional s-wave state without sign reversal mediated by orbital fluctuation.\cite{Kontani,Yanagi}

\begin{figure}[b]
\begin{center}
\includegraphics[width=8cm]{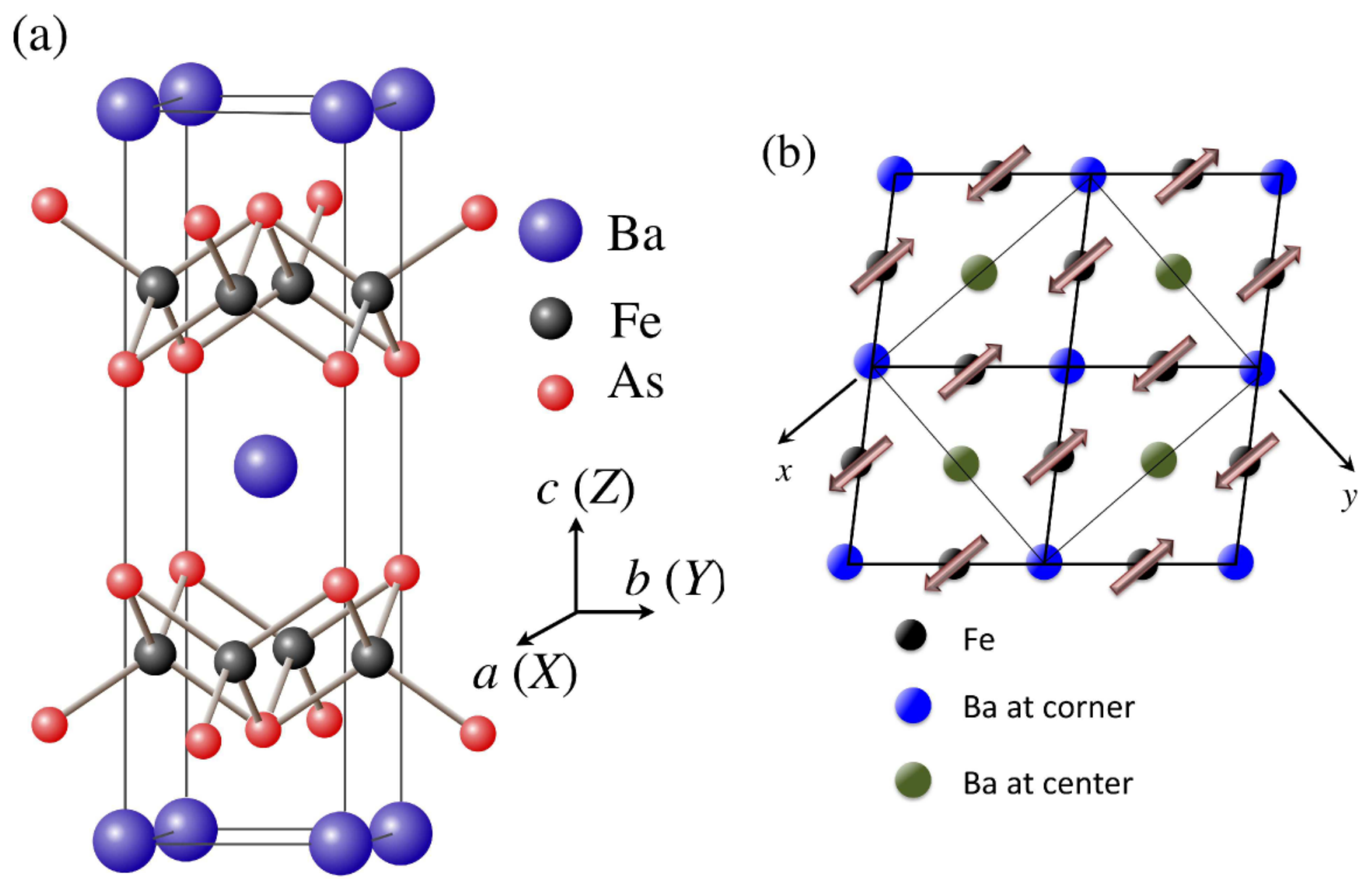}
\end{center}
\caption{(Color online) (a) Crystal structure of BaFe$_{2}$As$_{2}$, which belongs to the base-centered tetragonal crystal class $I4/mmm$, and (b) crystal structure and magnetic order below $T_{\rm S}$ with the crystal symmetry of $Fmmm$. 
Figure cited from M. Yoshizawa et al., J. Phys. Soc. Jpn. {\bf 81} (2012) 024604.
\copyright 2012, Journal of the Physical Society of Japan.}
\label{f1}
\end{figure}

\begin{figure*}[t]
\begin{center}
\includegraphics[width=15cm]{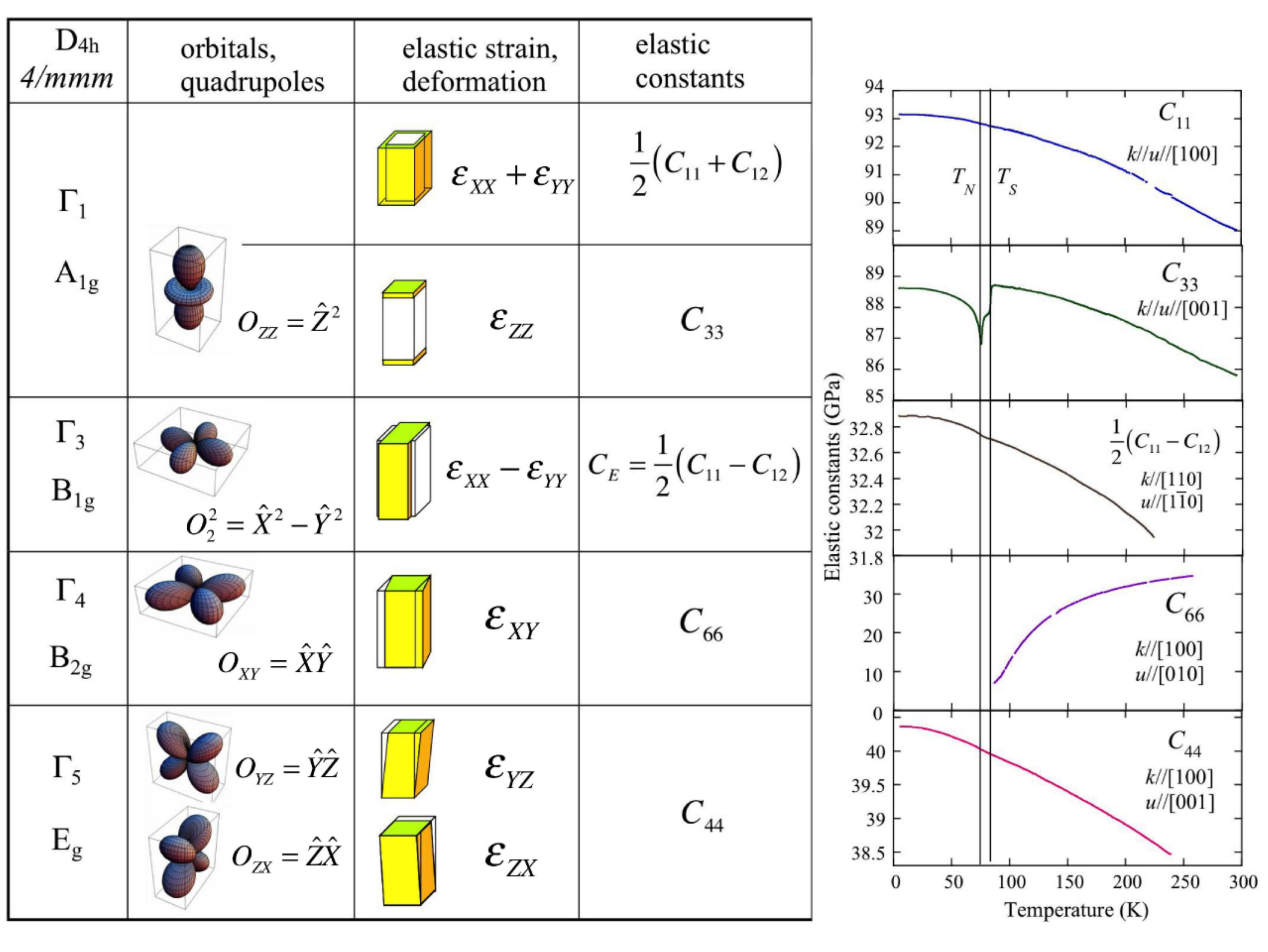}
\end{center}
\caption{(Color online) Orbitals, quadrupoles, elastic strains, and their corresponding elastic constants classified into the irreducible representation of the point group $D_{4h}$. Temperature dependence
of all elastic constants of Ba(Fe$_{0.963}$Co$_{0.037}$)$_{2}$As$_{2}$ are shown in the figure. 
Figure cited from M. Yoshizawa et al., Mod. Phys. Lett. B {\bf 26} (2012) 1230011. 
\copyright 2012, Modern Physics Letters B.}
\label{f2}
\end{figure*}

Figure \ref{f1} shows the crystal structure of BaFe$_{2}$As$_{2}$, which is tetragonal at room temperature. 
It becomes orthorhombic at a structural transition temperature of $T_{\rm S}=$140 K, accompanied by the appearance of a long-range antiferromagnetic order\cite{Rotter}.
By replacing Fe with Co, $T_{\rm S}$ tends to decrease, and superconductivity appears.
On the elastic properties of iron-based superconductors, McGuire et al. reported a large elastic anomaly towards low temperatures for the 1111 system LaFeAsO$_{1-x}$F$_{x}$.\cite{McGuire}
Similar elastic anomalies have been reported for Co doped BaFe$_{2}$As$_{2}$ and BaFe$_{1.84}$Co$_{0.16}$As$_{2}$.\cite{Fernandes}
Large elastic softening has also been reported for the elastic constant $C_{\rm 66}$ of Ba(Fe$_{0.9}$Co$_{0.1}$)$_{2}$As$_{2}$.\cite{Goto} 
All experimental results indicated a considerable anomaly at the structural phase transition temperature. 
More recently, our group has reported a large elastic anomaly at the structural phase transition of Ba(Fe$_{1-x}$Co$_{x}$)$_{2}$As$_{2}$, where $x$ = 0, 0.037, 0.060, 0.084, 0.098, 0.116, 0.161, and 0.245, and revealed the roles of orbital fluctuations in the large elastic softening of $C_{66}$.\cite{Yoshizawa} These remarkable anomalies are consistent with the theoretical calculations, suggesting an important role of structural and orbital fluctuations for the emergence of superconductivity.\cite{KontaniSSC}

In previous studies, we have reported a structural fluctuation in $C_{\rm 66}$ associated with the structural and magnetic phase transitions of the iron-based superconductor Ba(Fe$_{1-x}$Co$_{x}$)$_{2}$As$_{2}$. 
The structural fluctuation gives rise to an in-plane order and then it has a two-dimensional nature. 
However, ordering hardly occurs in two-dimensional systems, so a three-dimensionality is necessary for the occurrence of ordering. 
In this work, we will focus on other elastic constants of Ba(Fe$_{1-x}$Co$_{x}$)$_{2}$As$_{2}$.
In particular, we will focus on to the role of the three-dimensional effect though the precise measurements of the elastic constant $C_{\rm 33}$.
Our experimental tool, ultrasonic measurement, provides us with information on the changes in the symmetry of the lattice system. 
Strains introduced into the crystal in the elastic constant measurements deform the crystal locally and break the crystal symmetry (symmetry breaking field).
The corresponding elastic strain $\varepsilon_{\rm ZZ}$ for $C_{\rm 33}$ does not change the crystal symmetry, but modulates inter-layer spacing. 
Then, we get information on the inter-layer spacing and effect of three-dimensionality.  
In this report, we will show a peculiar behavior of the $C_{\rm 33}$ of the iron-based superconductor Ba(Fe$_{1-x}$Co$_{x}$)$_{2}$As$_{2}$.

\section{Experimental Procedure}
Single-crystalline samples of Ba(Fe$_{1-x}$Co$_{x}$)$_{2}$As$_{2}$ were grown by the self-flux method. 
The size of the samples are typically 2 $\times$ 2 $\times$ 2 mm$^{3}$, which have rectangle shape to prevent various sound signals.
The Co concentration in the grown crystals was determined by EDS. 
Elastic constants were measured by the ultrasonic pulse-echo phase comparison method as a function of temperature from 5 to 300 K using a cryostat equipped on a Gifford-McMahon (GM) cryocooler. 
In this research, we report the results of the elastic constants of Ba(Fe$_{1-x}$Co$_{x}$)$_{2}$As$_{2}$, which has a tetragonal symmetry at room temperature, therefore it has six independent elastic constants such as $C_{\rm 11}$, $C_{\rm 33}$, $C_{\rm 12}$, $C_{\rm 13}$, $C_{\rm 44}$, $C_{\rm 66}$ respectively. It is difficult to measure $C_{\rm 13}$ using a conventional experimental configuration, so we measured the other elastic constants of this sample by choosing the propagation direction and the displacement of the sound wave, as shown in Fig. \ref{f2}. In ultrasonic measurement, sound traveling in a solid deforms the crystals locally, lowering the symmetry of the sample. 
The strains introduced into the solid were classified into irreducible representations of the point group, as shown in Fig. \ref{f2}. 
In this measurement, ultrasound was emitted and detected using LiNbO$_{3}$ transducers. To choose the electromagnetic signal, the operating fundamental frequency was changed to one of its odd harmonics of the transducers. 
$Z$-cut LiNbO$_{3}$ with 100 ${\mu}$m thickness was used for longitudinal ultrasonic waves, and a 41$^\circ$ $X$-cut plate of LiNbO$_{3}$  with 100 ${\mu}$m thickness was used for transverse waves.
The fundamental frequencies of the longitudinal and transverse transducers were 33 and 19 MHz, respectively.
We usually used third-higher harmonics of 114 and 64 MHz to generate longitudinal and transverse sound waves, respectively. 
We measured the ultrasonic attenuation at a frequency of 38 MHz.
The elastic polymer Thiokol LP-32 was used to glue the transducers on both sides of the sample with parallel end faces. 
To prevent damage to the sample owing to the rapid changes in temperature, the rate of the change in temperature was controlled to be less than 10 K/h near the phase transitions.
Elastic constants were obtained using  $C = \rho v^{2}$, where $\rho$ is the density and $v$ is either the longitudinal or transverse sound velocity. 

\begin{figure}[t]
\begin{center}
\includegraphics[width=8cm]{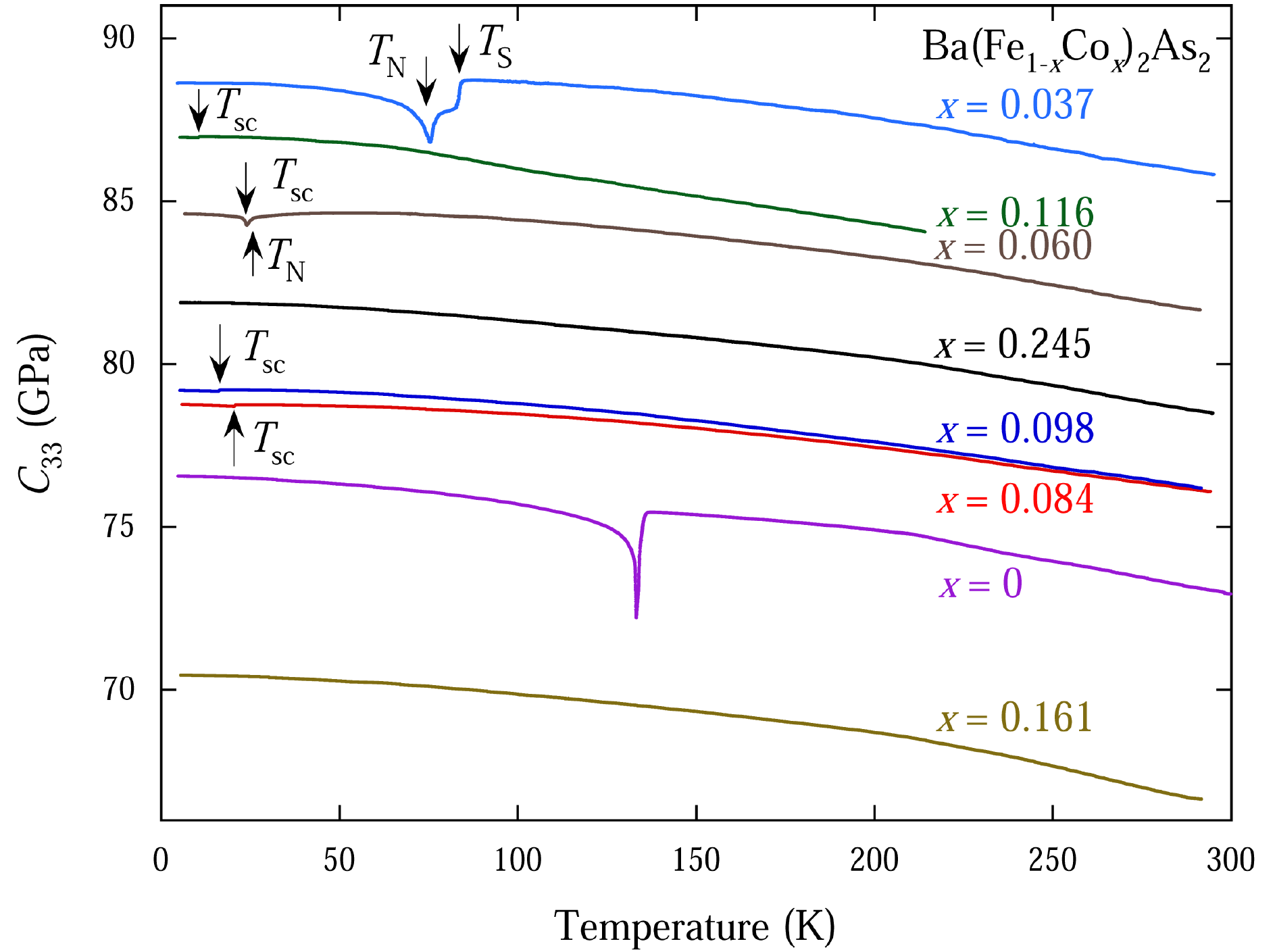}
\end{center}
\caption{(Color online) Temperature dependence of elastic constant $C_{\rm 33}$ of Ba(Fe$_{1-x}$Co$_{x}$)$_{2}$As$_{2}$ with various Co concentrations.}
\label{f3}
\end{figure}

\section{Results}
The longitudinal elastic constants $C_{\rm 11}$ and $C_{\rm 33}$, and the shear elastic constants $C_{\rm 44}$, $\frac{1}{2}$($C_{\rm 11}$ - $C_{\rm 12}$), and $C_{\rm 66}$ of Ba(Fe$_{0.963}$Co$_{0.037}$)$_{2}$As$_{2}$ are shown in Fig. \ref{f2}. All these elastic constants except for $C_{\rm 66}$ show a monotonic increase with decreasing temperature. 
We discussed the large elastic softening of $C_{\rm 66}$ in the previous report.\cite{Yoshizawa}
More interesting point we would like to discuss here is the observation of two remarkable anomalies in the elastic constant $C_{\rm 33}$ of Ba($_{0.963}$Co$_{0.037}$)$_{2}$As$_{2}$, which shows elastic anomalies associated with both structural and magnetic transitions, while the other elastic constants do not show any elastic anomaly at the magnetic phase transition. 

To investigate the elastic anomaly in $C_{\rm 33}$, we have carried out ultrasonic measurements for the samples with various Co concentrations.
Figure \ref{f3} shows the temperature dependences of the elastic constant $C_{\rm 33}$ of the Ba(Fe$_{1-x}$Co$_{x}$)$_{2}$As$_{2}$ samples with $x$ = 0, 0.037, 0.06, 0.084, 0.098, 0.116, 0.161, and 0.245. 
The samples with $x$ = 0, and 0.037 show elastic anomalies at $T_{\rm N}$ and $T_{\rm S}$.
The over-doped samples with $x$ = 0.084, 0.098, and 0.116 show a step-wise elastic anomaly at $T_{\rm sc}$.
The nearly optimally doped sample with $x$ = 0.060 shows remarkable elastic anomalies at $T_{\rm sc}$ and $T_{\rm N}$. 
The appearance of the elastic anomaly in $C_{\rm 33}$ for the under-doped, optimally doped, and over-doped samples indicates the three-dimensional character of the iron-based superconductor Ba(Fe$_{1-x}$Co$_{x}$)$_{2}$As$_{2}$. 

\begin{figure}[t]
\begin{center}
\includegraphics[width=8cm]{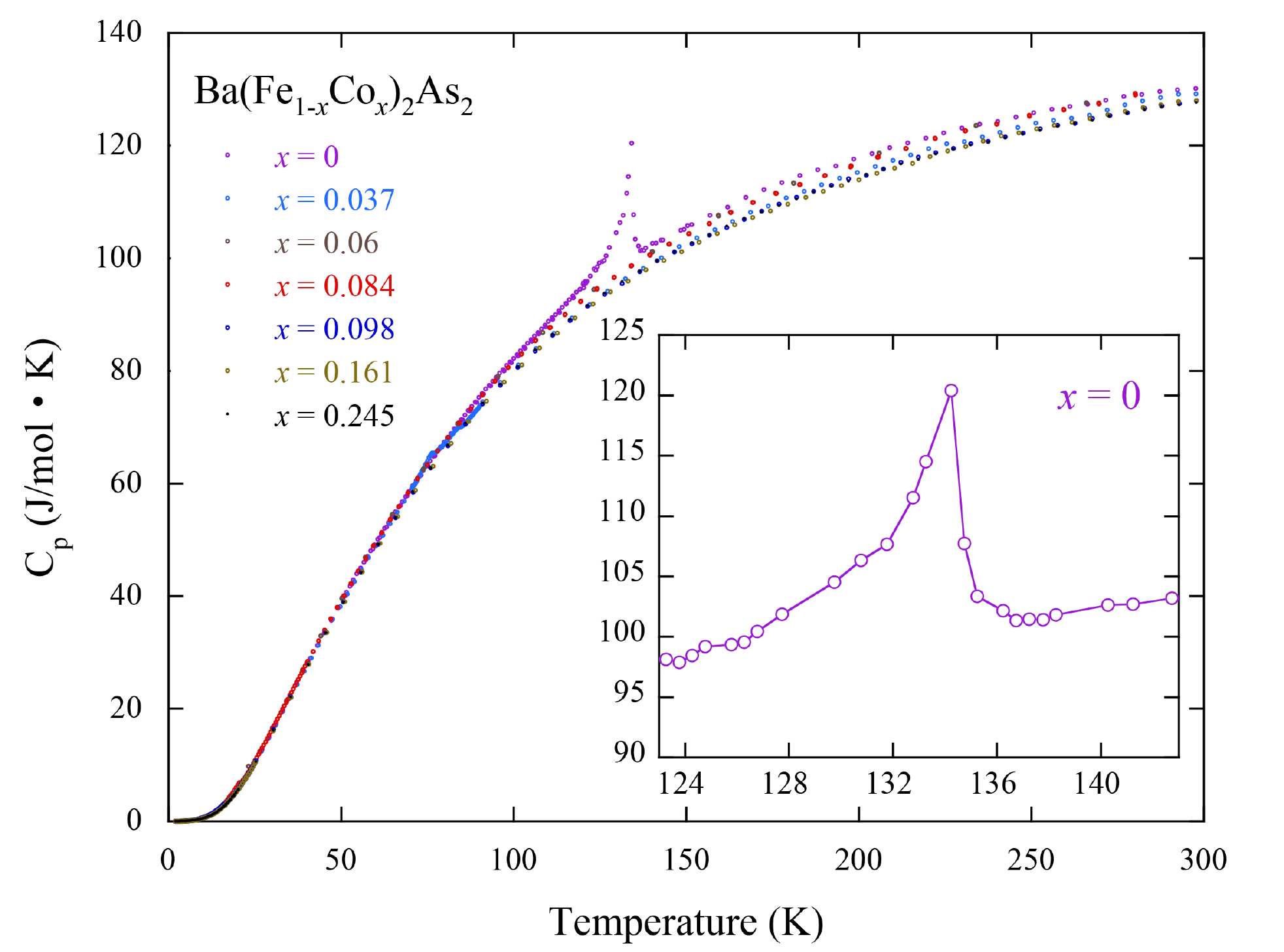}
\end{center}
\caption{(Color online) Heat capacity data for seven representative Co concentrations ($x$ = 0, 0.037, 0.060, 0.084, 0.098, 0.161, and 0.245). }
\label{f4}
\end{figure}

\begin{figure}[t]
\begin{center}
\includegraphics[width=8cm]{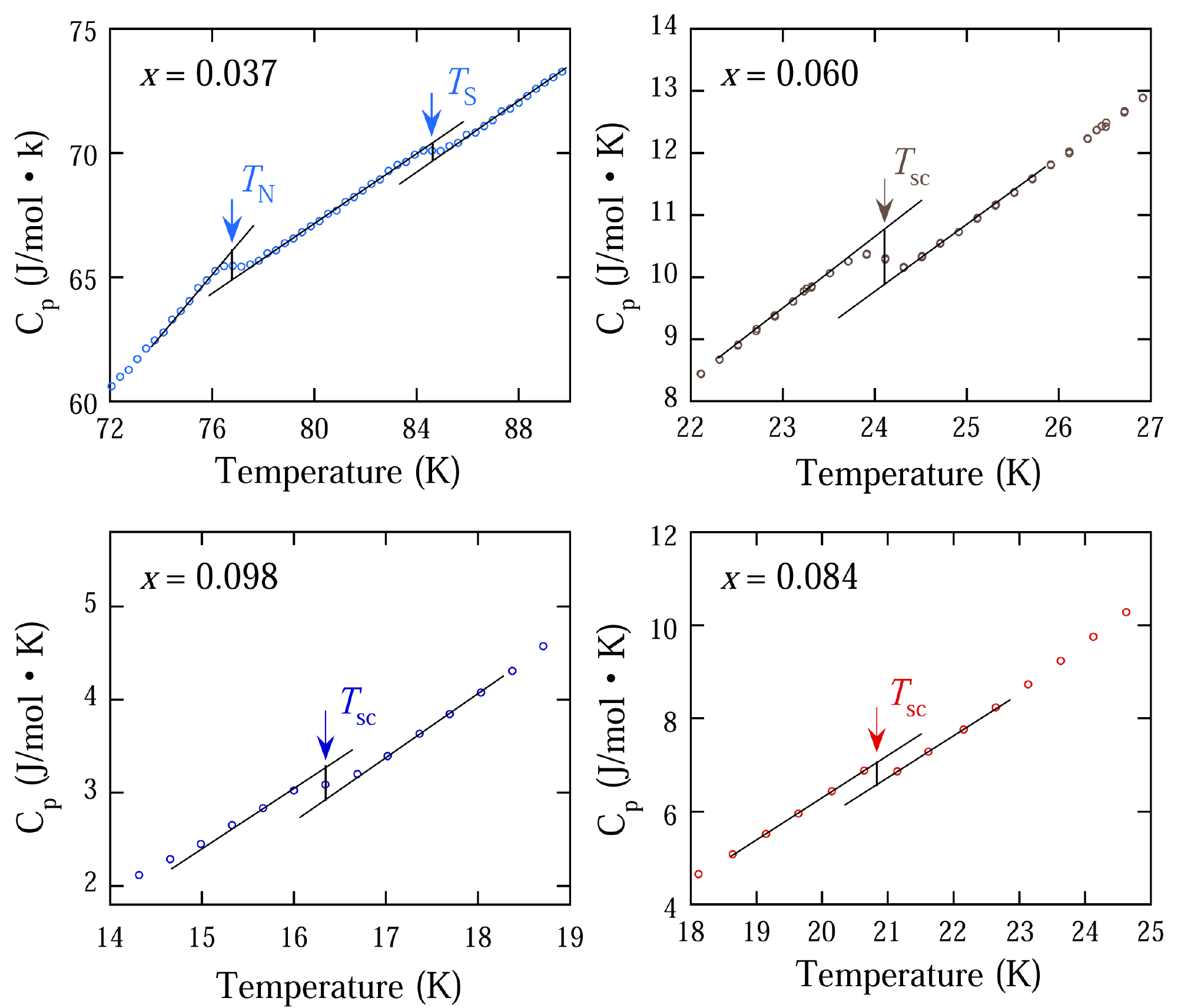}
\end{center}
\caption{(Color online) Heat capacity jump at the phase transition temperatures for $x$ = 0.037, 0.060, 0.084, and 0.098.}
\label{f5}
\end{figure}

Elastic anomaly was also observed in heat capacity, which was measured using a physical properties measurement system (PPMS) at each phase transition.
The results are shown in Fig. \ref{f5} for representative Co concentrations. 
We did not measure the heat capacity of the 11.6\% doped sample, because the sample weight exceeded the measurement range of the sample puck of PPMS. 
For the undoped BaFe$_{2}$As$_{2}$, a clear sharp peak was observed at the structural/magnetic phase transition, as shown in the inset of Fig. \ref{f4}. We can compare this result quantitatively with those of other studies.\cite{Budko, Jiun-Haw}
Figure \ref{f5} shows the heat capacity data near the transition temperatures for the sample with $x$ = 0.037, 0.060, 0.084, and 0.098. 
Our data shows clear specific jumps at the phase transition temperatures. 
We evaluated the Gr\"{u}neisen parameter using the elastic constant and heat capacity jumps at $T_{\rm sc}$. We will discuss this in detail later.

\begin{figure*}
\begin{center}
\includegraphics[width=15cm]{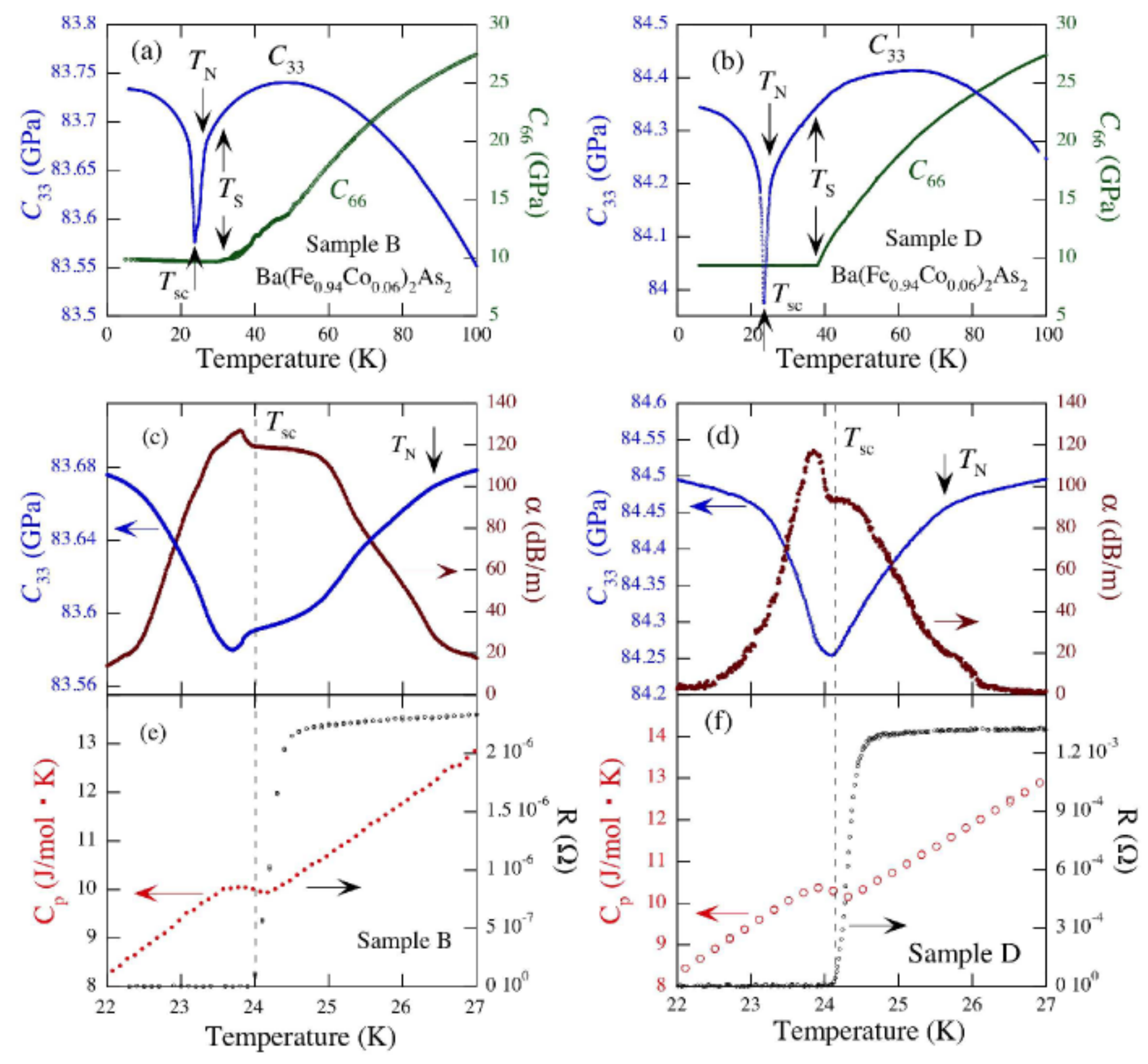}
\end{center}
\caption{(Color online) (a) Temperature dependence of elastic constant $C_{\rm 33}$ and $C_{\rm 66}$ for optimally doped sample B.  (b) Temperature dependence of elastic constant $C_{\rm 33}$ and $C_{\rm 66}$ for optimally doped sample D. (c) Temperature dependence of elastic constant $C_{\rm 33}$ and attenuation of longitudinal elastic constant for optimally doped sample B. (d) Temperature dependence of elastic constant $C_{\rm 33}$ and attenuation of longitudinal elastic constant for optimally doped sample D. (e) Temperature dependence of heat capacity and electrical resistivity near the superconducting phase transition temperature for optimally doped sample B. (f) Temperature dependence of heat capacity and electrical resistivity near the superconducting phase transition temperature for optimal doped sample D.
(In the figures: $T_{\rm sc}$ indicates the superconducting phase transition temperature, $T_{\rm S}$ indicates the structural phase transition temperature, and $T_{\rm N}$ indicates the magnetic phase transition temperature). The longitudinal attenuation in Fig. (c) and (d) was calculated using $\alpha\ $ = - $\frac{20}{vT}$log$\mid$A$\mid$, where $v$ is the longitudinal sound velocity, $T$ is the echo position, A is the echo strength.}
\label{f6}
\end{figure*}

Next, we will report the peculiar physical properties of Ba(Fe$_{0.94}$Co$_{0.06}$)$_{2}$As$_{2}$, which is located near the QCP. 
For this Co concentration, we have studied the elastic properties, specific heat, and electrical resistance of two samples from different batches. In this paper, we call them sample B and sample D. 
Although EDS measurement shows the same Co concentration for both samples, we suppose that the Co concentrations are different from each other. Precisely, we suppose that sample D has a higher Co concentration than sample B. 
A clear sample dependence for this composition was proved by careful inspection of the elastic constants $C_{\rm 33}$ and $C_{\rm 66}$, resistance, and elastic attenuation $\alpha$. 
According to the phase diagram of Ba122, the structural, magnetic, and superconducting phase transitions are closely located to each other for the 6.0\% doped sample.
To determine at which phases elastic anomalies are occurred, we carried out simultaneous measurements of the electrical resistance and elastic constant in one setting. We could not obtain resistivity owing to experimental limitations. Superconductivity is evident from the sharp drop in the resistance to zero of sample B at 24 K, 24.2 K for sample D.
These results of the electrical resistance associated with the superconducting transition were also followed by the heat capacity measurement. 

Figures \ref{f6}(a) and (b) show the temperature dependence of the elastic constants $C_{\rm 33}$ and $C_{\rm 66}$.
$C_{\rm 66}$ shows large elastic softening from room temperature to low temperatures, and an inflection at $T_{\rm S}$. 
In the case of $C_{\rm 66}$, both samples show elastic softening at $T_{\rm S}$, but a small elastic anomaly was observed at $T_{\rm sc}$ and no anomaly at $T_{\rm N}$.
In the case of $C_{\rm 33}$, note that both samples show no anomaly at $T_{\rm S}$. However, three characteristic points attracted our attention: (1) The elastic constant $C_{\rm 33}$ gradually decreases from 50 - 60 K; we call this the elastic softening from the analogy of a large elastic anomaly in $C_{\rm 66}$. (2) The elastic constant $C_{\rm 33}$ drops steeply at $T_{\rm N}$. (3) Sample B shows a step-wise anomaly at $T_{\rm sc}$, but sample D shows no such anomaly at $T_{\rm sc}$.

According to the specific heat and resistance in Figs. \ref{f6}(e) and (f), a specific heat anomaly was observed at $T_{\rm sc}$, but not at $T_{\rm N}$.
The elastic constant $C_{\rm 33}$ of sample B and sample D show a rather large softening towards low temperatures. 
The amounts of softening are 0.2\% for sample B and 0.5\% for sample D. 
For this analysis, it would be interesting to determine the main factor for the softening, which starts from the high-temperature region such as 50 - 60 K.
In general, it has been reported that fluctuations associated with superconductivity do not start from such a high temperature, and appear just above $T_{\rm sc}$. 
They cannot induce large elastic softening. 
To the best of our knowledge, the largest elastic softening associated with superconductivity appears in the organic superconductor $\kappa$-(ET)$_{2}$X (X = Cu(NCS)$_{2}$, Cu[N(CN)$_{2}$]Br), which was discovered by Simizu et al.\cite{Simizu}
Even in this case, the softening starts from 1.2$T_{\rm sc}$ at most. 
On the other hand, the softening of $C_{\rm 33}$ starts near the 2$T_{\rm sc}$ for sample B and 3$T_{\rm sc}$ for sample D, implying that the elastic anomaly in $C_{\rm 33}$ is not from the superconducting origin. 
From our results, we would like to conclude that the anomaly in $C_{\rm 33}$ is possibly ascribed to magnetic origin, although the origin still remains unknown. 

We have also studied ultrasonic attenuation $\alpha$ of the longitudinal elastic waves propagating along the $c$-axis for both samples.
In Figs. \ref{f6}(c), and (d), $C_{\rm 33}$ and $\alpha$ are depicted as functions of temperature on an expanded scale.
Although there are precise differences between sample B and D, we found interesting common features in $C_{\rm 33}$ and $\alpha$.
$\alpha$ shows a maximum at around $T_{\rm sc}$ and an additional peak below $T_{\rm sc}$ for both samples.
A similar maximum in ultrasonic attenuation was reported for BaFe$_{1.85}$Ni$_{0.15}$As$_{2}$ polycrystalline.\cite{Saint-paul}
Generally, ultrasonic attenuation decreases below $T_{\rm sc}$.
Historically, some heavy-electron superconductors, UBe$_{13}$ and UPt$_{\rm 3}$ showed an ultrasonic attenuation peak below $T_{\rm sc}$, which was discussed theoretically on the basis of Landau-Khalatnikov damping mechanism.\cite{Golding,Mueller,Miyake}
This mechanism is associated with the relaxation of the order parameter amplitude, which was observed in superfluid He at first.\cite{Williams}
From the same point of view as that in heavy fermion superconductors, the coupling between the superconducting gap amplitude and long wave length phonons was discussed for the case of a charge density-wave (CDW) compound NbSe$_{2}$.\cite{Littlewood}
The attenuation peak below $T_{\rm sc}$ for Ba(Fe$_{0.94}$Co$_{0.06}$)$_{2}$As$_{2}$ would be caused by the same origin as the heavy-fermion superconductors and /or CDW compound, however the origin is still open.

In elastic constant measurements, there are two types of coupling between the order parameter $M$ and the elastic strain $\varepsilon$: bilinear coupling with the form $M \varepsilon$, and magneto-elastic coupling with the form $M ^{2}\varepsilon$. Bilinear coupling brings about large elastic softening from high temperatures, which was seen in $C_{66}$. On the other hand, magneto-elastic coupling brings about only a step-wise elastic anomaly. 
In general, it does not show a large anomaly compared with bilinear coupling. In 6.0\% doped sample, the softening in $C_{33}$ starts from high temperatures, which indicates that the large softening in $C_{33}$ is due to bilinear coupling. 
It would also be a possible origin of the $C_{33}$ anomaly, but it is an enigma. Important point is that there is no anomaly at $T_{\rm S}$, but there is an anomaly at $T_{N}$. 
This suggests the importance of magneto-elastic coupling in the $C_{33}$ elastic anomaly. For the over-doped samples, $C_{\rm 33}$ shows a small but sharp softening towards $T_{\rm sc}$, so the elastic anomaly in $C_{\rm 33}$ for the sample with $x$ = 0.060, 0.084, 0.098, and 0.116 suggests that the coupling between the order parameter and elastic strain is $M ^{2}\varepsilon$.

\section{Discussion}
\begin{figure}[b]
\begin{center}
\includegraphics[width=8cm]{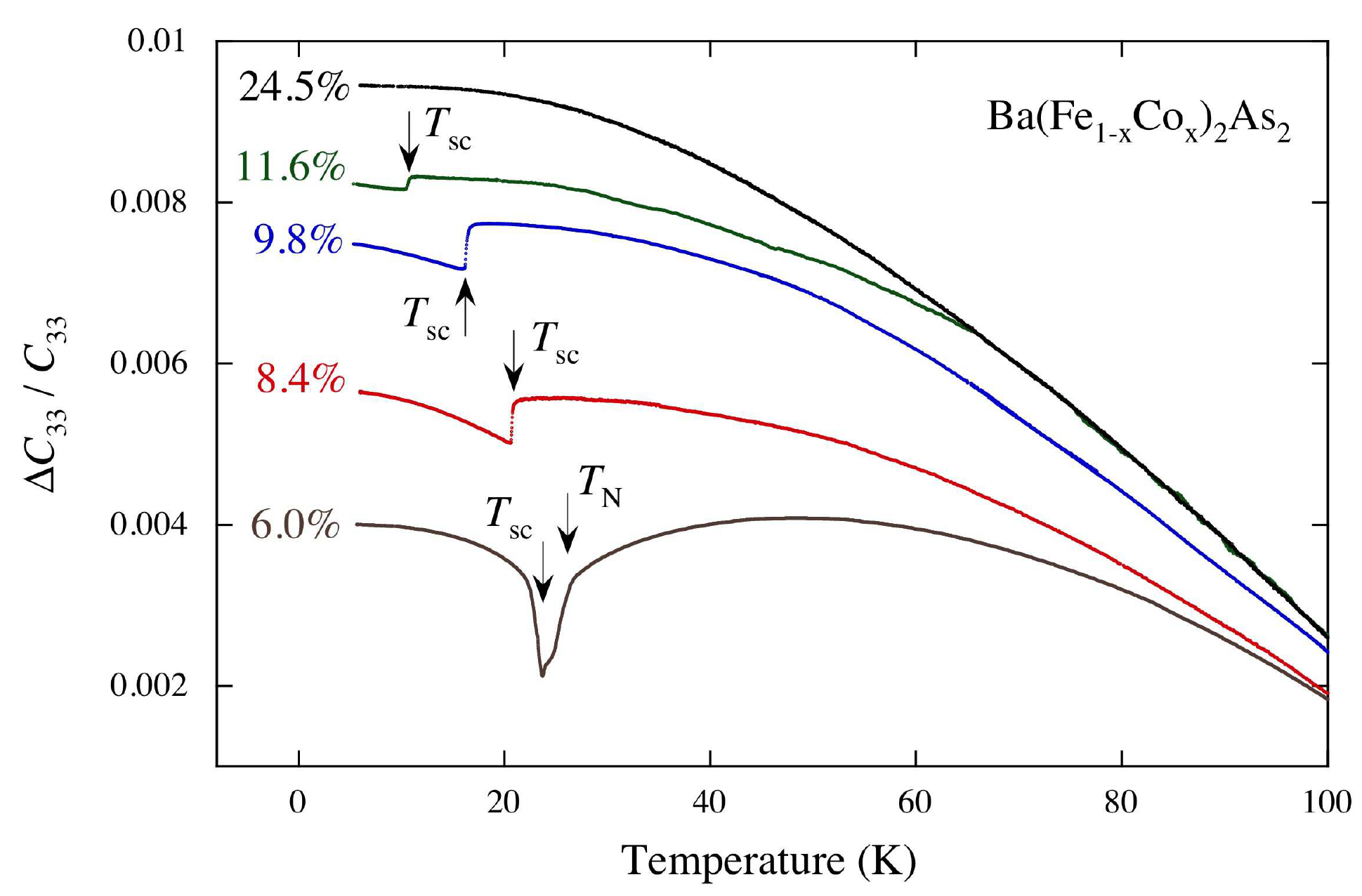}
\end{center}
\caption{(Color online) Normalized elastic constant $\Delta C_{\rm 33}/C_{\rm 33}$ of optimally doped and over-doped samples, over-doped sample 24.5\% describes background elastic constant. }
\label{f7}
\end{figure}
\begin{figure}[t]
\begin{center}
\includegraphics[width=8cm]{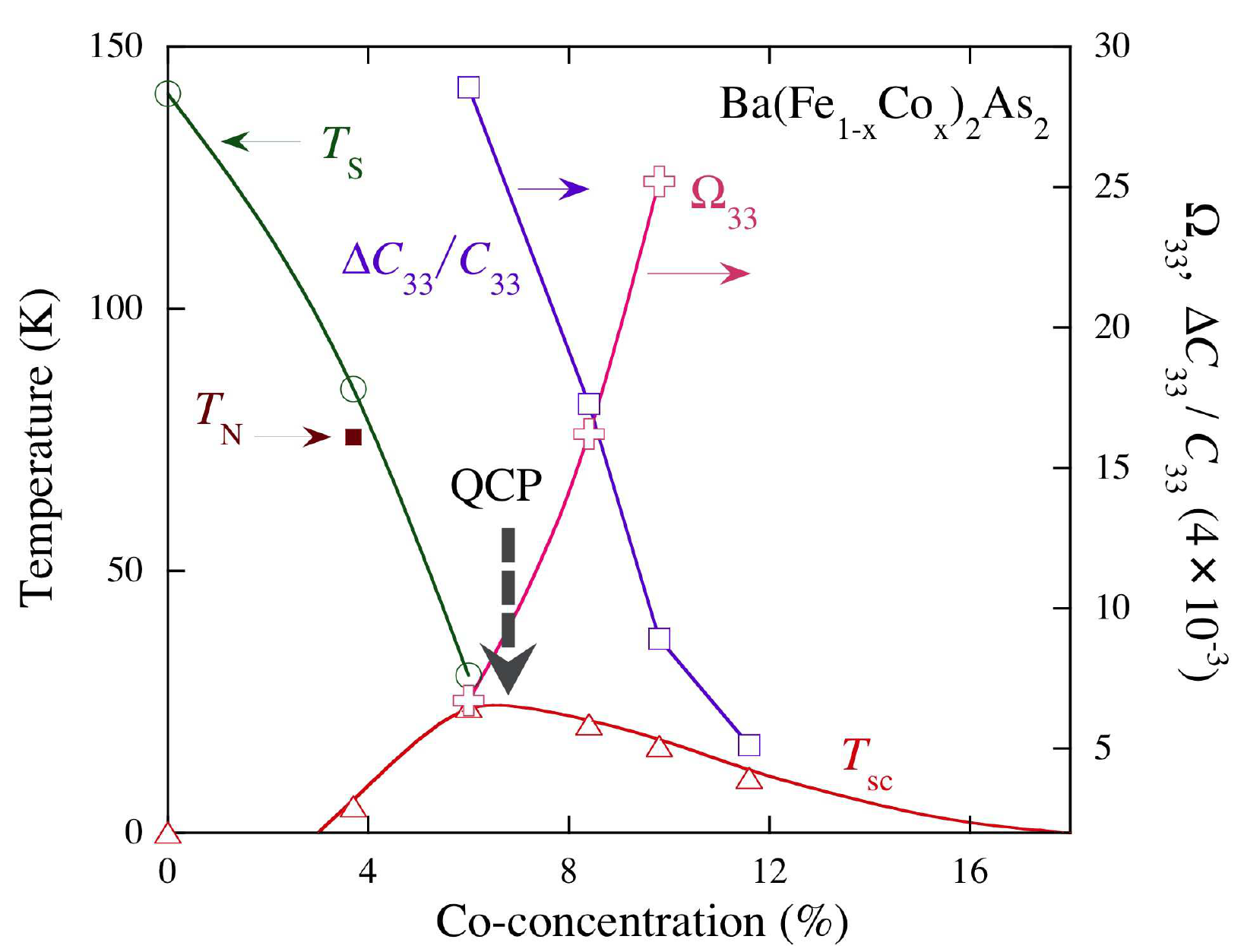}
\end{center}
\caption{(Color online) Phase diagram of Ba(Fe$_{1-x}$Co$_{x}$)$_{2}$As$_{2}$. Gr{\" u}neisen constant calculated using Eq. (1), and the amount of softening $C_{\rm 33}-C_{\rm 33,0}$ estimated from Fig. \ref{f7}.}
\label{f8}
\end{figure}
In previous studies, we discussed the correlation of $T_{\rm sc}$ with $S_{\rm 66}$. 
Namely, structural fluctuations participate in the emergence of superconductivity. 
This means that in-plane two-dimensional fluctuation mediates the pairing mechanism in iron-based superconductors. 
On the other hand, according to the Wigner-Marmin theory, ordering is hardly occurs in a two-dimensional system. 
Three-dimensionality is necessary for the occurrence of the ordering. 
In this paper, we have reported the enhancement of the inter-layer fluctuation near the QCP. 
On Ba(Fe$_{1-x}$Co$_{x}$)$_{2}$As$_{2}$, we have shown that $T_{\rm sc}$ correlates with $S_{\rm 66}$ (1/$C_{\rm 66}$).
Namely, superconductivity is enhanced with lattice softening.

In Fig. \ref{f7}, we plotted the relative amount of softening, which is normalized the data at 120 K as 
${{\Delta {C_{\rm 33}}} \mathord{\left/
 {\vphantom {{\Delta {C_{\rm 33}}} {{C_{\rm 33}}}}} \right.
 \kern-\nulldelimiterspace} {{C_{\rm 33}}}}\left( {x,T} \right) = {{{C_{\rm 33}}\left( {24.5 \%,T} \right)} \mathord{\left/
 {\vphantom {{{C_{\rm 33}}\left( {24.5 \%,T} \right)} {{C_{\rm 33}}\left( {x,T} \right)}}} \right.
 \kern-\nulldelimiterspace} {{C_{\rm 33}}\left( {x,T} \right)}} - 1$.
The normalized data $\Delta C_{\rm 33}/C_{\rm 33}$ merge at 120 K. The overdoped samples show a remarkable step-wise anomaly at $T_{\rm sc}$. Unlike the 6.0\% doped sample, above $T_{\rm sc}$ in overall temperature range, the overdoped samples show monotonic increase with decreasing temperature. However, when we regard the 24.5\% sample as the background, the overdoped samples show an anomaly towards low temperatures same as the 6.0\% doped sample. The amount of softening is defined by 
$\Delta C_{33}/C_{33}=\Delta C_{33}(24.5 \%, T_{\rm sc})/C_{33}(24.5 \%, 120 K)-\Delta C_{33}( x, T_{sc}(x))/C_{33}(x, 120 K)$.
This amount becomes larger as approach to the QCP, as shown in Fig. \ref{f8}.
As in the relation between $T_{\rm sc}$ and $S_{\rm 66}$, $C_{\rm 33}$ softening is considered to be correlated with $T_{\rm sc}$.
In other words, the inter-layer fluctuation component brings about high $T_{\rm sc}$ in agreement with the in-plane fluctuation appearing in $S_{\rm 66}$.
Most surprisingly, three-dimensionality possibly originated from the magnetic character of these materials.

Second, we will discuss the Gr{\" u}neisen parameter, which is a scale of magnitude and order parameter, and the so-called interaction of the order parameter and distortion of the magneto-strictive coupling. 
We have determined the Gr{\"u}neisen parameter $\Omega_{\rm sc}$ for $T_{\rm sc}$ from in jump in the specific heat and the jump in the elastic modulus at $T_{\rm sc}$ by

\begin{equation}
 \Delta C=-\Omega^{2}_{\rm sc}\Delta C_{\rm v}T_{\rm sc}.
\end{equation}

Figure \ref{f8} shows the Co concentration dependence of $\Omega_{\rm sc}$.
Gr{\"u}neisen parameter has a small value near the QCP, and gradually increases with increasing of Co concentration in the over-doped region. 
Our estimation can be checked from the results in previous works.
The Gr{\"u}neisen parameter is defined as 

\begin{equation}
\Omega_{\rm sc} = - \frac{1}{T_{\rm sc}}\frac{dT_{\rm sc}}{d\varepsilon_{\rm i}}.
\end{equation}

\noindent Here, the uniaxial strain dependence $dT_{\rm sc}/d\varepsilon_{\rm i}$ is related to the uniaxial pressure dependence of $T_{\rm sc}$ as

\begin{align}
\frac{dT_{\rm sc}}{d\varepsilon_{\rm xx}}&=-C_{\rm 11}\frac{dT_{\rm sc}}{dp_{\rm a}}-C_{\rm 12}\frac{dT_{\rm sc}}{dp_{\rm a}}-C_{\rm 13}\frac{dT_{\rm sc}}{dp_{\rm c}} .\\
\frac{dT_{\rm sc}}{d\varepsilon_{\rm zz}}&=-C_{\rm 33}\frac{dT_{\rm sc}}{dp_{\rm c}}-2C_{\rm 13}\frac{dT_{\rm sc}}{dp_{\rm a}} .
\end{align}

\noindent where $i$ is 1 for $XX$ and the $a$-axis, and 3 for $ZZ$ and the $c$-axis.
$C_{\rm ij}$ is the corresponding elastic constants, and ${dT_{\rm sc}}/{dp_{\rm i}}$ is the uniaxial pressure dependence of $T_{\rm sc}$.
We can calculate $\Omega$ from the elastic constant and uniaxial pressure dependence of $T_{\rm sc}$.

We would like to compare our results with these of previous works.
Bud$^{'}$ko et al. reported $dT_{\rm sc}/dp_{\rm i}$ for 3.8\% and 7.4\%, and Hardy et al. reported $dT_{\rm sc}/dp_{\rm i}$ for 8.0\%. 
Bud$^{'}$ko et al. obtained $dT_{\rm sc}/dp_{\rm a}=-4.1$ K/kbar and $dT_{\rm sc}/dp_{\rm c}=1.7$ K/kbar for 3.8\% doped sample, $dT_{\rm sc}/dp_{\rm a}=0.3$ K/kbar and $dT_{\rm sc}/dp_{\rm c}=-2.6$ K/kbar for 7.4\% doped sample \cite{Budko}. 
Hardy et al. obtained $dT_{\rm sc}/dp_{\rm a}=3.1(1)$ K/GPa and $dT_{\rm sc}/dp_{\rm c}=-7.0(2)$ K/GPa for 8\% doped sample \cite{Hardy} from the thermal expansion measurement.
We used $C_{\rm 11}$, $C_{\rm 33}$, and $C_{\rm 12}$ values of 109.2, 78.7, and 43.46 GPa, respectively, for the calculation.
Since $C_{\rm 13}$ cannot be obtained by our measurements, we assumed it to be the same value as $C_{\rm 12}$.
In the case of 8\% doped sample, the Gr\"{u}neisen constant and $dT_{\rm sc}/d\varepsilon_{\rm ZZ}$ are evaluated to be 14.2 and 282 K, respectively.
The value of $\Omega$ is consistent with our result.
On the other hand, the predicted values did not achieved for the 3.8\% and 7.4\% doped samples, they were ten times larger than our value. 
For the 3.8\% doped sample, the calculated values are 716 K for $dT_{\rm sc}/d\varepsilon_{\rm ZZ}$ and 102.3 for the Gr\"{u}neisen constant, they are 2037 K and 97 for the 7.4\% doped sample.
These results are inconsistent with our results and Hardy et al.'s
The reason for the differences remains unknown. 

Here, we have to pay attention to the sign of the Gr{\" u}neisen parameter.
We cannot determine the information about whether $\Omega$ is positive or negative, when it is evaluated from Eq. (1).
On the other hand, Eqs. (2) - (4) give its sign.
$dT_{\rm sc}/dp_{\rm c}$ is positive for underdoped samples, and negative for the overdoped doped samples.
$dT_{\rm sc}/dp_{\rm a}$ has an opposite sign of $dT_{\rm sc}/dp_{\rm c}$. 
Since the 6.0\% doped sample is located in the underdoped region, the sign of $\Omega$ might be negative.
In this research, the parameter $dT_{\rm sc}/dp_{c}$ obtained by Gr\"{u}neisen calculation is positive for the underdoped samples and negative for the overdoped samples. If we consider the hydrostatic pressure dependence of this system, the change of $T_{\rm sc}$ is positive for the underdoped samples and negative for the overdoped samples. Negative change of $T_{\rm sc}$ for an overdoped sample was reported by Nakashima et al.\cite{Nakashima} These experimental results suggest that the hydrostatic pressure dependence of $T_{\rm sc}$ mainly originated from the c-axis property.

The calculated results of both $dT_{\rm sc}/d\varepsilon_{\rm a}$ and $dT_{\rm sc}/d\varepsilon_{\rm c}$ as a functions of Co concentration dependence are listed in Table I for the overdoped samples. 
Figure \ref{f8} shows a summary of the phase diagram of Ba(Fe$_{1-x}$Co$_{x}$)$_{2}$As$_{2}$. 
We found two characteristic parameters: Gr\"{u}neisen constant $\Omega$ and $\Delta C_{\rm 33}/C_{\rm 33}$. The Gr\"{u}neisen constant becomes small near the QCP, at which the superconducting transition temperature $T_{\rm sc}$ takes a maximum value, and the amount of softening in $C_{\rm 33}$ also takes a maximum value. This implies a relationship between the Gr\"{u}neisen constant and the inter-layer fluctuations near the QCP.

\begin{table*}
\caption{$T_{\rm sc}$, $\Delta C_{\rm 33}$, $\Delta C_{\rm p}/T_{\rm sc}$, $dT_{\rm sc}/dp_{\rm c}$, calculated $dT_{\rm sc}/{d\varepsilon_{\rm zz}}$, and  $\Omega$ values for the 6.0\%-, 8.4\%-, and 9.8\%-doped samples.}
\centering
\begin{tabular}{p{0.12\linewidth}p{0.1\linewidth}p{0.1\linewidth}p{0.1\linewidth}p{0.1\linewidth}p{0.1\linewidth}p{0.1\linewidth}p{0.1\linewidth}p{0.1\linewidth}p{0.1\linewidth}}
\hline
x-Co(\%)    & $T_{\rm sc}$ (K) & $\Delta C_{\rm 33}$ (10$^{-2}$GPa)  & $\Delta C_{\rm p}/T_{\rm sc}$ (mJ/mol$\cdot K^{2}$)  & $dT_{\rm sc}/{dp_{\rm c}}$ K/GPa & $dT_{\rm sc}/{d\varepsilon_{\rm zz}}$ &  $\mid\Omega\mid$ \\
\hline
6          & 24 & 1.4 & 33 & 1.9 & 160 & 6.7   \\
\hline
8.4         & 20.6 & 4.2 & 23 & 4.2 &  334 & 16.2  \\
\hline
9.8      & 16.7 & 4.4 & 15 & 5.3 &  420 & 25.2   \\
\hline
\end{tabular}
\end{table*}

\section{Conclusions}
In this work, we focused our attention on the role of the three-dimensional effect in the emergence of superconductivity. 
We reported the characteristic temperature dependence of the elastic constant $C_{\rm 33}$ of Ba(Fe$_{1-x}$Co$_{x}$)$_{2}$As$_{2}$. 
In particular, $C_{\rm 33}$ shows elastic anomalies near the QCP, where a large elastic softening starts from high temperatures.
We have studied it by measuring specific heat, resistivity, and ultrasonic attenuation in addition to the elastic constant.
The anomaly of $C_{\rm 33}$ is due to the enhancement of inter-layer fluctuation near QCP.  
These data show that the three-dimensional properties of the Ba122 system appearing in $C_{\rm 33}$ would be related to magnetic fluctuation.
The amount of inter-layer fluctuation is correlates with $T_{\rm sc}$ for the overdoped samples.

According to Drotzinger et al. the phase diagram of Ba(Fe$_{1-x}$Co$_{x}$)$_{2}$As$_{2}$ is nicely corresponds to that under hydrostatic pressure $p$. 
They looked for a key parameter, which is commonly affected by $x$ and $p$, and reported that the most likely candidate is the Fe-As distance, because this quantity shows similar $x$ and $p$ dependences.\cite{Drotzinger}
The Fe-As distance will be modulated by in-plane distortion and inter-layer compression.
This is the most likely reason why the amount of inter-layer fluctuation appearing in $C_{\rm 33}$ correlates with $T_{\rm sc}$ as well as the in-plane fluctuation appearing in $C_{\rm 66}$.
Lee et al. proposed the so-called Lee Plot for the $T_{\rm sc}$ of the iron-based superconductors, where it is implied that a higher $T_{\rm sc}$ appears as being close to the right tetrahedral structure.\cite{Lee}
This conjecture has been studied on the basis of both orbital theory and spin fluctuation theory.\cite{Saito,Usui,Kuroki2}
The correlation between local structure and superconductivity also supported the importance of three-dimensional nature for the emergence of superconductivity.
Applying a pressure along the $c$-axis also mediates the bond angles of As-Fe-As, and reproduces magnetic order. 
The results of the ultrasonic investigation of the iron-based superconductors suggests both in-plane fluctuation and inter-layer fluctuation.
In-plane fluctuation accompanying an enormous lattice fluctuation stops at $T_{\rm S}$.

In this work, we found inter-layer fluctuation accompanying a small lattice fluctuation.
It does not stop below $T_{\rm S}$ and continues up to $T_{\rm N}$.
We obtained the Gr\"{u}neisen parameter along the $c$-axis. 
It becomes small near the QCP.
This is very crucial information, because it is a reflection of 
${{d{T_{{\rm{sc}}}}} \mathord{\left/
 {\vphantom {{d{T_{{\rm{sc}}}}} {d{\varepsilon _{ZZ}}}}} \right.
 \kern-\nulldelimiterspace} {d{\varepsilon _{ZZ}}}}$, 
 which can be obtained only by elastic constant measurement.
Generally, ${{d{T_{{\rm{sc}}}}} \mathord{\left/
 {\vphantom {{d{T_{{\rm{sc}}}}} {d{p_c}}}} \right.
 \kern-\nulldelimiterspace} {d{p_c}}}$ can be measured by
for the uniaxial pressure measurement.
However, this quantity contains the effect of the compression of the $c$-axis and the expansion of $ab$-plane.
Therefore, it is not clear which is relevant, $c$-axis compression or $ab$-plane expansion, from the uniaxial pressure experiment along $c$-axis.
Our data, which is consistent with Hardy et al.'s result, suggests that controlling the $c$-axis length is very important in the realization of high $T_{\rm sc}$ in iron-based superconductors.
Recent works have discussed the high $T_{\rm sc}$ by controlling the $c-$axis length in the superconductivity for iron-based superconductors. \cite{Kazutaka, Sunagawa}
The inter-plane effect reported in this paper is as important as the in-plane effect appearing in $C_{\rm 66}$.
We suppose that both fluctuations correlate with $T_{\rm sc}$, and would  participate collaboratively to the emergence of superconductivity. 
We hope that our experimental results will spotlight the influence of uniaxial pressure effects and the role and relation of orbital and magnetic fluctuations in the emergence of superconductivity, and will stimulate further experimental and theoretical studies.

\section*{Acknowledgments}
The authors wish to thank H. Fukuyama, H. Kontani, and Y. {\^ O}no for valuable discussions, and T. Kowata, R. Kamiya, R. Onodera for assistance in the experiments.
This work was supported by a Grant-in-Aid for Scientific Research on Innovative Areas, "Heavy Electrons" (No. 20102007), by a Grant-in-Aid for Scientific Research on Challenging Exploratory Research (No. 22654037) from the Ministry of Education, Culture, Sports, Science and by Technology, Japan, and by the Transformative Research Project on Iron Pnictides of the Japan Science and Technology Agency.

\end{document}